\documentclass[twocolumn,showpacs,floats,floatfix,superscriptaddress,aps,prl]{revtex4}
\usepackage{amsfonts}
\usepackage{amssymb}
\usepackage{amsmath}
\usepackage{graphicx}
\usepackage{bm}

\begin{document}

\author{Jonathan Busch}
\affiliation{The School of Physics and Astronomy, University of Leeds, Leeds LS2 9JT, United Kingdom}
\author{Elica S. Kyoseva}
\affiliation{The School of Physics and Astronomy, University of Leeds, Leeds LS2 9JT, United Kingdom}
\affiliation{Department of Physics, Sofia University, James Bourchier 5 blvd, 1164 Sofia, Bulgaria}
\author{Michael Trupke}
\affiliation{Blackett Laboratory, Imperial College London, Prince Consort Road, London SW7 2BZ, United Kingdom}
\author{Almut Beige}
\affiliation{The School of Physics and Astronomy, University of Leeds, Leeds LS2 9JT, United Kingdom}

\title{Entangling distant quantum dots using classical interference}
\date{\today}

\begin{abstract}
We show that it is possible to employ \emph{reservoir engineering} to turn two distant and relatively bad cavities into one good cavity with a tunable spontaneous decay rate. As a result, quantum computing schemes, that would otherwise require the shuttling of atomic qubits in and out of an optical resonator, can now be applied to {\em distant} quantum dots. To illustrate this we transform a recent proposal  to entangle two qubits via the observation of macroscopic fluorescence signals [Metz {\em et al.}, Phys. Rev. Lett. {\bf 97}, 040503 (2006)] to the electron-spin states of two semiconductor quantum dots. Our scheme requires neither the coherent control of qubit-qubit interactions nor the detection of single photons. Moreover, the scheme is relatively robust against spin-bath couplings, parameter fluctuations, and the spontaneous emission of photons.
\end{abstract}

\maketitle

The strength of measurement-based quantum computing is that its performance
is independent of the experimental parameters. Whenever a
certain measurement outcome is obtained, the system is projected onto a well-defined state with a very high fidelity. This is useful when the final state is highly entangled or
differs from the initial one only by a desired quantum gate operation. One
example of measurement-based quantum computing are linear optics schemes based on the detection of single photons \cite{linopt}. Further examples are the processing of atomic qubits via the detection of single or no photons \cite%
{Cabrillo} and the manipulation of the electron-spin states of
quantum dots via charge detection \cite{Beenakker}. However, the scalability of these approaches depends
strongly on the respective measurement efficiency.

The entangling scheme by Metz \emph{et al.} \cite{Metz} alleviates the detection problem via the observation of macroscopic quantum jumps \cite{Dehmelt}. This means, the interactions in the system are engineered such
that it emits a random telegraph signal of long periods of intense fluorescence (light periods) interrupted by long periods of no photon emission (dark periods). The successful state preparation is heralded by a macroscopic dark period. Ref.~\cite{Metz} describes a scheme that prepares two laser driven atoms inside an optical cavity in a maximally entangled state. The same authors have shown that electron shelving techniques allow even for the build up of large cluster states \cite{Metz2}. However, this requires the shuttling  of atomic qubits in and out of an optical cavity, which is time consuming and susceptible to decoherence.

\begin{figure}[t]
\begin{minipage}{\columnwidth}
\begin{center}
\resizebox{\columnwidth}{!}{\rotatebox{0}{\includegraphics{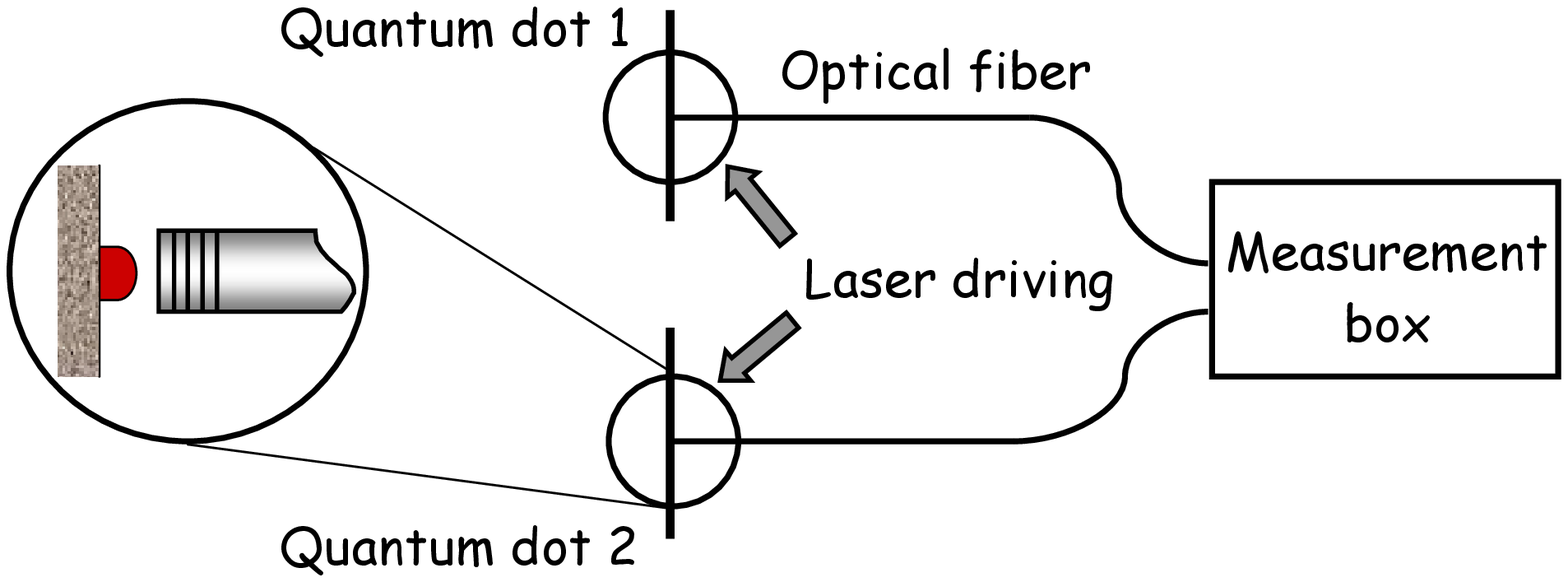}}}
\end{center}
\vspace*{-0.5cm} \caption{Experimental setup to entangle two distant quantum dots via the observation of macroscopic quantum jumps.}
\label{scheme}
\end{minipage}
\end{figure}

In this letter, we propose a scheme for distributed entanglement preparation with inherent scalability. We require neither the transport of qubits from one interaction zone to another nor the detection of single photons. This is achieved via reservoir engineering based on an interference effect that has already been observed experimentally \cite{Eichmann,Grangier}. For example, in their famous two-atom double-slit experiment, Eichmann \emph{et al.} \cite{Eichmann} detected the spontaneously emitted photons from two laser-driven ions on a distant photographic plate. The pattern formed resembles that of a classical double-slit experiment. In fact, the ions behave like classical dipole emitters sending out electromagnetic waves \cite{Schon}. Here we use optical fibers to create an analogous coupling of two distant cavities to common modes of the free radiation field. 

Qubits placed in these distant cavities experience only one common resonator mode, a so-called bus mode \cite{spiller}, with a tunable spontaneous decay rate. Consequently, it is now possible to generalize ideas for the generation of scalable entanglement in atom-cavity systems to solid state systems. As in previous quantum dot schemes (see e.g.~Refs.~\cite{prop}), we encode information in electron-spin states. Each dot is driven by a laser field and placed inside an optical cavity. This is feasible with current technology \cite{cav}. The light coming from the fibers is constantly monitored by a distant measurement box (c.f.~Fig.~\ref{scheme}). 

The detected fluorescence signal exhibits macroscopic quantum jumps such that a dark period indicates the shelving of the qubits in a maximally entangled state. Transitions from one fluorescence period into another are now caused by spin-bath couplings, parameter fluctuations, or the spontaneous emission of photons into free space. These jumps play a vital role in the proposed state preparation and make it relatively robust against experimental imperfections. We require only that the cavities experience the same system-bath interaction. The quantum dots do not have to be identical.

\begin{figure}[t]
\begin{minipage}{\columnwidth}
\begin{center}
\resizebox{\columnwidth}{!}{\rotatebox{0}{\includegraphics{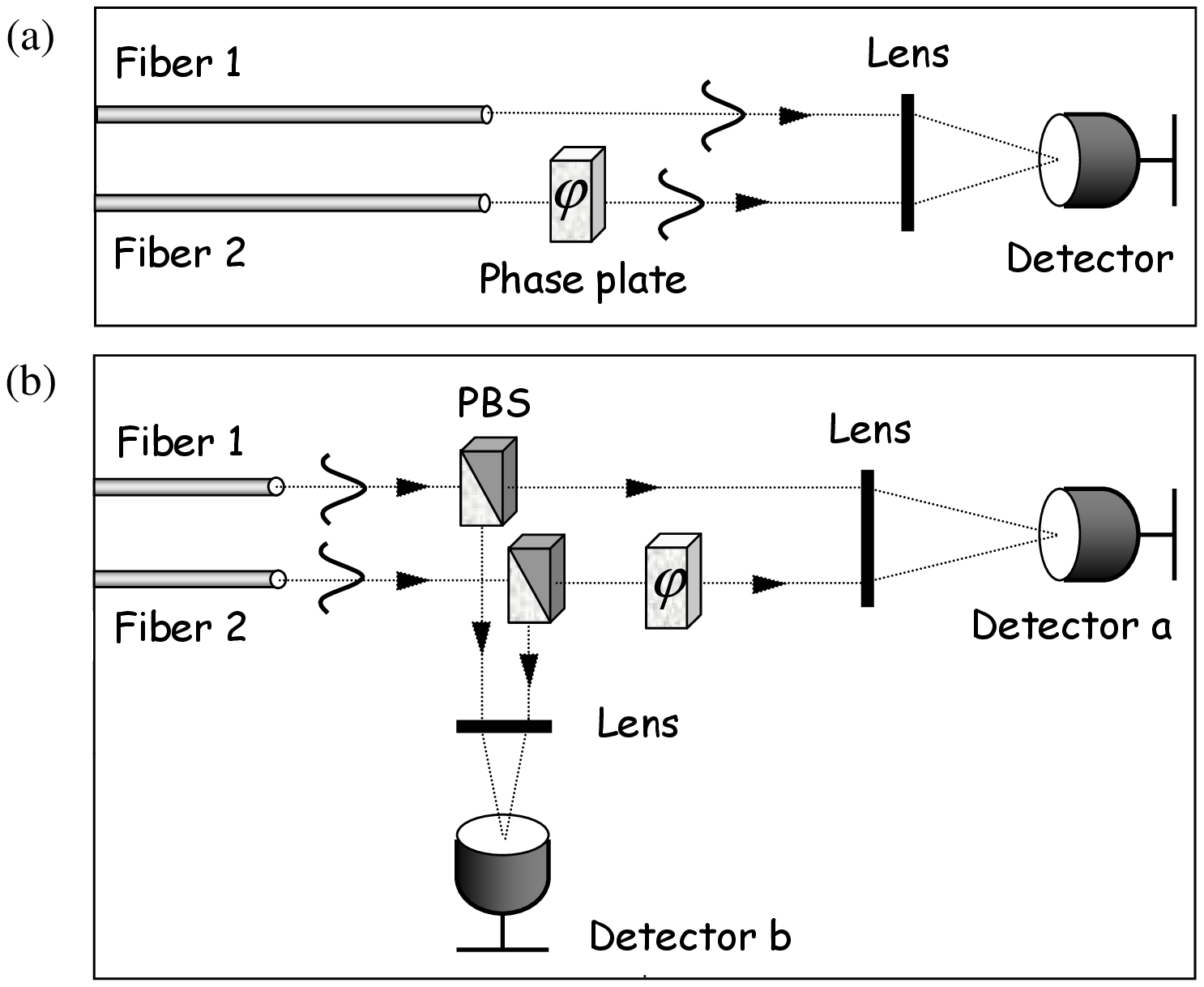}}}
\end{center}
\vspace*{-0.5cm} \caption{Measurement box in Fig.~\ref{scheme}. (a) Illustration of the interference of cavity fiber photons. (b) Box to entangle two qubits using, in addition, two polarising beam splitters (PBS's), a second photon detector and two lenses.}\label{fig4}
\end{minipage}
\end{figure}

In the following, spontaneous emission due to the interaction between the cavity field modes and the radiation field propagating through the optical fibers 1 and 2 is taken into account by the dipole Hamiltonian 
\begin{equation} 
H_{\mathrm{dip}}=\sum_{i=1,2}e\,\mathbf{D}_{i}\cdot \mathbf{E}(\mathbf{r}%
_{i})\,.  \label{Hdip}
\end{equation}
Here $\mathbf{D}_{i}$ is an effective dipole moment and proportional to the sum of the annihilation and the creation operator $c_{i}+c_{i}^{\dagger}$ for a photon in cavity $i$ and $\mathbf{E}(\mathbf{r}_{i})$ describes the radiation field, where cavity $i$ couples to fiber $i$. Suppose $a_{\mathbf{k}}^{\dagger }$ is the creation operator for a photon with wave vector $\mathbf{k}$ and frequency $\omega _{k}$ and both cavities have the same fiber coupling constants $g_{\mathbf{k}}$, the same frequency $\omega _{\mathrm{c}}$, and the same linear polarisation. Then $H_{\rm dip}$ in Eq.~(\ref{Hdip}) equals, in the rotating wave approximation, 
\begin{eqnarray} \label{Hdip2}
H_{\mathrm{dip}} &=&\sum_{\mathbf{k}} \hbar g_{\mathbf{k}} \left[ \mathrm{e%
}^{\mathrm{i}\mathbf{k}\cdot \mathbf{r}_{1}}\,c_{1}+ \mathrm{e}^{\mathrm{i}(\mathbf{k}\cdot \mathbf{r}_{2}+\varphi )}\,c_{2}\right] a_{\mathbf{k}}^{\dag } + \mathrm{H.c.} ~~~~~
\end{eqnarray}%
for the measurement box depicted in Fig.~\ref{fig4}(a). Since the fiber photons are constantly monitored, a quantum jump description \cite{Plenio,Schon} based on Eq.~(\ref{Hdip2}) can be used to show that the conditional Hamiltonian for the no-photon time evolution of the system equals
\begin{equation}
H_{\mathrm{cond}}= H_{\mathrm{cond}}' - {{\frac{\mathrm{i}}{2}}}\hbar \kappa_{\rm c} \,c(\varphi)^{\dagger }c(\varphi) \label{Hcond} \, . 
\end{equation}
Here $H_{\mathrm{cond}} '$ is the no-photon time evolution Hamiltonian in the absence of cavity decay, $c(\varphi)$ is the annihilation operator 
\begin{equation}
c(\varphi) \equiv \left[ \,\mathrm{e}^{\mathrm{i}\mathbf{k}_{c}\cdot \mathbf{r}%
_{1}}\,c_{1}+\mathrm{e}^{\mathrm{i}(\mathbf{k}_{c}\cdot \mathbf{r}%
_{2}+\varphi) }\,c_{2}\,\right] /\sqrt{2} \, , \label{ca}
\end{equation}%
$\kappa_{\rm c}$ is the spontaneous decay rate of the common cavity mode $c(\varphi)$, and $\mathbf{k}_{\mathrm{c}}$ is the fiber photon wave vector. In case of a cavity photon emission, the state of the system changes from $|\psi \rangle $ into $c(\varphi)|\psi \rangle /\Vert c(\varphi)|\psi \rangle \Vert $.

Notice that the two cavities possess two common cavity modes (one with annihilation operator $c(\varphi)$ and one with $c(\varphi + \pi)$) but only one decay channel. Photons in the $c(\varphi + \pi)$ mode cannot leak out through the cavity mirrors and have a negligible decay rate. Their electromagnetic field components interfere destructively, thereby making the detection of $c(\varphi + \pi)$ photons in the fiber impossible. Field components created by $c(\varphi)$ photons, to the contrary, interfere constructively and can be detected easily. Consequently, the two cavities do not emit photons independently but are effectively coupled via the radiation field in the fibers. The role of the lens in Fig.~\ref{fig4}(a) is to maximize this coupling by focussing the light from the different cavities onto the same spot on the detector, thereby safely erasing all information about the origin of the arriving photons. 

Let us now have a look at a variation of the above setup, in which the measurement box in Fig.~\ref{fig4}(a) is replaced by the measurement box shown in Fig.~\ref{fig4}(b). The new box contains a second photon detector and two parallel polarising beam splitters. Depending on their orientation, some parts of the light field in the fibers are now reflected towards detector $b$. In analogy to Eq.~(\ref{Hdip2}), the interaction of the cavity photons with the environment is given by the Hamiltonian 
\begin{eqnarray}
H_{\mathrm{dip}} &=&\sum_{\mathbf{k}}\hbar g_{\mathbf{k};a} \left[ \mathrm{e%
}^{\mathrm{i}\mathbf{k}\cdot \mathbf{r}_{1}}\,c_{1}+ \mathrm{e}^{\mathrm{i}(\mathbf{k}\cdot \mathbf{r}_{2}+\varphi )}\,c_{2}\right] a_{\mathbf{k}}^{\dag }  \notag  \label{Hdip3} \\
&&+\hbar g_{\mathbf{k};b} \left[ \mathrm{e}^{\mathrm{i}\mathbf{k}\cdot 
\mathbf{r}_{1}}\,c_{1}+\mathrm{e}^{\mathrm{i}\mathbf{k}%
\cdot \mathbf{r}_{2}}\,c_{2}\right]b_{\mathbf{k}}^{\dag }+\mathrm{H.c.}~~~
\end{eqnarray}%
Here the $b_{\mathbf{k}}^{\dagger }$ are the creation operators for the photons arriving at detector $b$ and the $g_{\mathbf{k};a}$ and the $g_{\mathbf{k};b}$ are fiber coupling constants. If we assume, for example, 
\begin{equation}
\varphi =\pi ~~\mathrm{and}~~\mathrm{e}^{\mathrm{i}\mathbf{k}_{\mathrm{c}%
}\cdot \mathbf{r}_{1}}=\mathrm{e}^{\mathrm{i}\mathbf{k}_{\mathrm{c}}\cdot 
\mathbf{r}_{2}}\,,  \label{max}
\end{equation}
then the no-photon time evolution of the system is described by the conditional Hamiltonian 
\begin{equation} \label{condi}
H_{\mathrm{cond}}=H_{\mathrm{cond}} ' -\sum_{x=a,b}{{\frac{\mathrm{i}}{2}}}\hbar
\kappa _{x}\,c_{x}^{\dagger }c_{x} 
\end{equation}%
with the annihilation operators 
\begin{equation}
c_a \equiv ( c_{1} - c_{2} )/\sqrt{2} ~~ {\rm and} ~~
c_b \equiv ( c_{1} + c_{2} )/\sqrt{2} \,.  \label{cb}
\end{equation}
Here $\kappa_a$ and $\kappa_b$ are the decay rates of the common cavity $a$ and $b$-modes. The ratio of both rates can be adjusted to any size by simply changing the orientation of the polarizing beam splitters. In the following we assume that 
$\kappa_a$ is much larger than $\kappa_b$ and all other relevant parameters. As we shall see below, the cavity $a$-mode is then so overdamped that it plays effectively no role in the time evolution of the system. Consequently, the two cavities operate like a single cavity with the {\em tunable} spontaneous decay rate $\kappa_b$!

\begin{figure}[t]
\begin{minipage}{\columnwidth}
\begin{center}
\resizebox{\columnwidth}{!}{\rotatebox{0}{\includegraphics{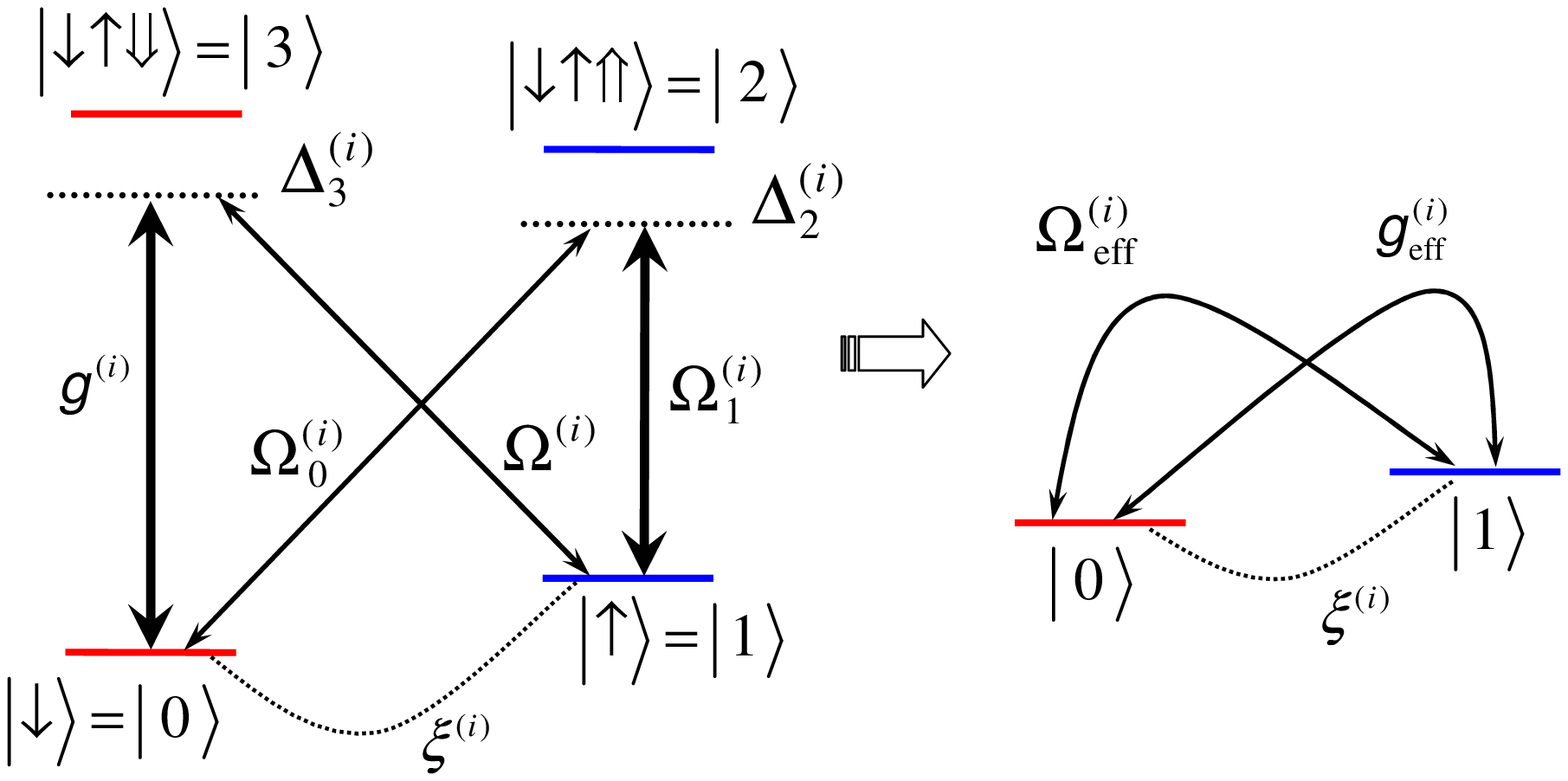}}}
\end{center}
\vspace*{-0.5cm} \caption{Level configuration and effective level scheme of a single quantum dot.} \label{level_scheme}
\end{minipage}
\end{figure}

Suppose now, each cavity is a part of a quantum dot-cavity system \cite{cav}. The internal level configuration of each quantum dot is shown in Fig.~\ref{level_scheme} and should be as in a recent experiment by Atat\"ure \emph{et al.} \cite{server} on spin-state preparation with near-unity fidelity. In the ground states $|0 \rangle = \left\vert \downarrow \right\rangle$ and $|1 \rangle = \left\vert \uparrow \right\rangle$, the dot contains one spin up or one spin down electron with
angular momentum projection $m_{z}=-1/2$ and $m_{z}=+1/2$. In the excited
states $\left\vert 2\right\rangle = \left\vert \downarrow \uparrow
\Downarrow \right\rangle $ and $\left\vert 3\right\rangle = \left\vert
\downarrow \uparrow \Uparrow \right\rangle$, the dot contains two electrons
in a singlet state and a heavy hole with spin projections $m_{z}=-3/2$ and $
m_{z}=+3/2$. The 1--2 dipole transition of dot $i$ is driven by a circularly
polarised laser with Rabi frequency $\Omega _{1}^{(i)}$ and detuning $\Delta_2^{(i)}$. Additional laser fields drive the quadrupole transitions 0--2 and 1--3 with
Rabi frequency $\Omega _{0}^{(i)}$ and $\Omega ^{(i)}$ and detuning 
$\Delta _{2}^{(i)}$ and $\Delta _{3}^{(i)}$. The 0--3 transition
couples with detuning $\Delta _{3}^{(i)}$ and coupling constant $g^{(i)}$ to the
quantised mode of cavity $i$. Introducing the rotating wave approximation and choosing the appropriate interaction picture, the Hamiltonian $H_{\rm cond} '$ in Eq.~(\ref{condi}) becomes time independent and equals 
\begin{eqnarray}
H_{\mathrm{cond}} ' &=&\sum_{i=1,2}\Big[\hbar g^{(i)}\,|0\rangle _{ii}\langle
3|\,c_{i}^{\dag }+\sum_{j=0,1}{{\frac{1}{2}}}\hbar \Omega
_{j}^{(i)}\,|j\rangle _{ii}\langle 2|  \notag  \label{hsys} \\
&&+{{\frac{1}{2}}}\hbar \Omega ^{(i)}\,|1\rangle _{ii}\langle 3|+ {\frac{1}{2}}\hbar \xi ^{(i)}\,|0\rangle _{ii}\langle 1| + \mathrm{H.c.}\notag \\
&& +\sum_{j=2,3}\hbar \Big( \Delta _{j}^{(i)} - {{\rm i} \over 2} \Gamma_j^{(i)} \Big) \,|j\rangle _{ii}\langle j| \Big]\,.
\end{eqnarray}%
Here the $\Gamma_j^{(i)}$ are the decay rates of the states $|2 \rangle$ and $|3 \rangle$. The $\xi^{(i)}$-dependent terms take uncontrolled spin-bath interactions into account which mix the states $|0\rangle $ and $|1\rangle$ \cite{xi}. Without restrictions, we can assume that $\Omega_{j}^{(i)}$,~$\Omega^{(i)}$, and $g^{(i)}$ are real by including their phases in the definition of $|2 \rangle$, $|3 \rangle$ and the cavity photon states.

In the following, we show that the population of the cavity $a$-mode and spontaneous decay of electron-hole pair states remains negligible when 
\begin{equation}
\Omega _{j}^{(i)}, \, \Omega ^{(i)}, \, g^{(i)} \, \sim \, \kappa _{b} \, \ll \, \Delta_{2}^{(i)}, \, \Delta _{3}^{(i)}, \, \kappa _{a} \, .
\end{equation}%
In this parameter regime, the $a$-mode and the states $|2 \rangle$ and $|3 \rangle$ can be eliminated adiabatically. Doing so, we find that photons in the $b$-mode effectively evolve on a much faster time scale than the qubit states $| 0 \rangle$ and $|1 \rangle$. Taking this into account and performing another adiabatic elimination, the Hamiltonian $H_{\rm cond}$ in Eq.~(\ref{condi}) simplifies~to  
\begin{eqnarray} \label{Hcond2}
H_{\mathrm{cond}} &=& \sum_{i=1,2}\Big[{{\frac{1}{2}}}\hbar \big(\xi ^{(i)} - \Omega _{
\mathrm{eff}}^{(i)} \big)\,|0\rangle _{ii}\langle 1| +\mathrm{H.c.} \notag \\
&& - \sum_{j=0,1}\hbar \Delta _{\mathrm{eff};j}^{(i)} \, |j\rangle _{ii}\langle
j| - \frac{\mathrm{i}}{2} \hbar \kappa_{\mathrm{eff}}^{(i,i)} \, |1\rangle _{ii}\langle 1| \Big] \notag \\
&& - {\frac{\mathrm{i}}{2}}\hbar \kappa _{\mathrm{eff}}^{(1,2)}\, \Big[ |10\rangle \langle 01| + |01 \rangle \langle 10| \Big]
\end{eqnarray}
with $\Omega _{\mathrm{eff}}^{(i)}\equiv \Omega_{0}^{(i)}\Omega _{1}^{(i)}/ 2\Delta _{2}^{(i)}$, $\Delta _{\mathrm{eff};0}^{(i)}\equiv \Omega _{1}^{(i)2} / 4\Delta_{2}^{(i)}$, $\Delta _{\mathrm{eff};1}^{(i)}\equiv \Omega_{1}^{(i)2} / 4\Delta _{2}^{(i)} + \Omega ^{(i)2} / 4\Delta _{3}^{(i)}$, and $\kappa _{\mathrm{eff}}^{(i,j)}\equiv \Omega^{(i)} \Omega^{(j)} g^{(i)} g^{(j)}/ 2 \kappa_b \Delta_{3}^{(i)} \Delta_{3}^{(j)}$ up to second order in $1/ \Delta_j$. The situation described by this Hamiltonian is analogous to the situation considered in Ref.~\cite{Metz}, where both qubits experience the same effective coupling constants, for small $\xi^{(i)}$ and when 
\begin{eqnarray} \label{conditions}
{\Omega_0^{(1)} \Omega_1^{(1)} \over \Omega_0^{(2)} \Omega_1^{(2)}} = {\Delta_2^{(1)} \over \Delta_2^{(2)}}\, , ~~
 {g^{(1)2} \over g^{(2)2}} = {\Omega^{(1)2} \over \Omega^{(2) 2}} = {\Delta_3^{(1)} \over \Delta_3^{(2)}} \, .
\end{eqnarray}
In this case, the Hamiltonian in Eq.~(\ref{Hcond2}) equals, up to an overall energy shift,  
\begin{eqnarray} \label{Hcond3} 
H_{\mathrm{cond}} &=& -{\frac{\hbar }{2\sqrt{2}}}\Big[ \Delta \xi \,\big(|00\rangle \langle a_{01}| - |a_{01}\rangle \langle 11|\big) \notag \\
&& \hspace*{-1.1cm} + \big(2\Omega _{\mathrm{eff}}^{(1)} -\xi ^{(1)}-\xi ^{(2)}\big) \big(|00\rangle \langle s_{01}| +|s_{01}\rangle \langle 11|\big)+ \mathrm{H.c.}\Big] \notag \\  
&& \hspace*{-1.1cm}  + \hbar \big( \Delta _{\mathrm{eff};0}^{(1)} - \Delta _{\mathrm{eff};1}^{(1)} \big) \big(|00\rangle \langle 00| - |11 \rangle \langle 11|\big)  \notag \\
&& \hspace*{-1.1cm} -\frac{\mathrm{i}}{2} \hbar \kappa _{\mathrm{eff}}^{(1,1)}\,\big(|s_{01}\rangle \langle s_{01}| + |11\rangle \langle{11}|\big)
\end{eqnarray}
with $\Delta \xi \equiv \xi ^{(1)}-\xi ^{(2)}$, $|a_{01}\rangle \equiv (|01\rangle -|10\rangle)/ \sqrt{2}$, and $|s_{01}\rangle \equiv (|01\rangle +|10\rangle )/\sqrt{2}$.

For $\Delta \xi = 0$, there are no transitions between the symmetric and antisymmetric subspace. Once in a symmetric state, the system emits photons towards detector $b $. However, when the qubits are in the only antisymmetric qubit state $|a_{01}\rangle $, \emph{no} photons arrive at this detector. The detector signal hence reveals information about the state of the quantum dots. The overall effect of this is the continuous projection of the qubits either onto the symmetric subspace or onto $|a_{01}\rangle$.

\begin{figure}[t]
\begin{minipage}{\columnwidth}
\begin{center}
\resizebox{\columnwidth}{!}{\rotatebox{0}{\includegraphics{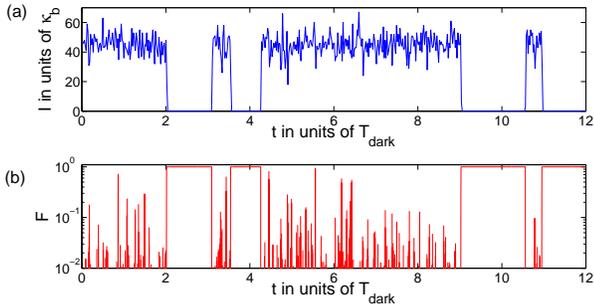}}}
\end{center}
\vspace*{-0.5cm} \caption{(a) Possible trajectory of the photon density $I(t)$ at detector $b$ obtained from a quantum jump simulation with $\Omega_{\rm eff}^{(1)} = \Delta_{\rm eff;0}^{(1)} = {1 \over 2} \Delta_{\rm eff;1}^{(1)} = {1 \over 4} \kappa_{\rm eff}^{(1,1)}$ and with $\xi^{(1)} = - \xi^{(2)} = 0.005 \, \kappa_b$. (b) Logarithmic plot of the corresponding fidelity $F$ of the maximally entangled state $|a_{01} \rangle$.}
\label{fig5}
\end{minipage}
\end{figure}

\begin{figure}[t]
\begin{minipage}{\columnwidth}
\begin{center}
\resizebox{\columnwidth}{!}{\rotatebox{0}{\includegraphics{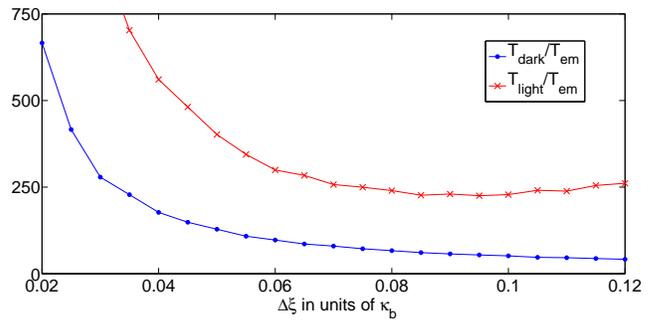}}}
\end{center}
\vspace*{-0.5cm} \caption{Quadratic dependence of the mean length of the light and dark periods, $T_{\rm dark}$ and $T_{\rm light}$, on $\Delta \xi$ obtained from a quantum jump simulation with $\Omega_{\rm eff}^{(1)} = \Delta_{\rm eff;0}^{(1)} = {1 \over 2} \Delta_{\rm eff;1}^{(1)} = {1 \over 4} \kappa_{\rm eff}^{(1,1)}$ and with $\xi^{(1)} = - \xi^{(2)}$.}
\label{fig6}
\end{minipage}
\end{figure}

In the case of small deviations $\Delta \xi \neq 0$, macroscopic quantum jumps occur from one subspace into the other. The system exhibits long periods of intense photon emissions (light periods) interrupted by long periods of no emission (dark periods) \cite{Dehmelt}, as shown in Fig.~\ref{fig5}(a). The population in $|a_{01}\rangle $ is very close to unity during a
dark period (c.f.~Fig.~\ref{fig5}(b)). A dark period hence prepares the qubits with a very high fidelity in a maximally entangled state. Identifying a successful state preparation is easy when the mean dark period length, $T_{\mathrm{dark}}$, is long compared to the mean time between photon emissions within a light period, $T_{\mathrm{em}}$. Due to the constant projection of the qubits, $T_{\mathrm{dark}}$ and  $T_{\mathrm{light}}$ scale as $1/\Delta \xi ^{2}$ (c.f.~Fig.~\ref{fig6}) and can be very long.

Transitions between light and dark periods are also caused by parameter
fluctuations violating condition (\ref{conditions}) and the spontaneous
emission of photons, which are not collected by the fibers. The effect of
these errors on the fidelity of the prepared state has already been studied in Ref.~\cite{Metz+} for an analogous setup. The analysis there suggests that spontaneous emission from excited states can be
tolerated, even if the system is operated in the vicinity of the bad-cavity
limit. Moreover, random variations of the coupling constants up to $30 \, \%$ do
not affect the fidelity of the prepared state. They only reduce the occurrence of relatively long dark periods.

\emph{In summary}, we have shown that it is possible to entangle distant quantum dots with electron-spin qubits via the observation of a macroscopic fluorescence signal. There are two conditions for the scheme to work. Eq.~(\ref{max}) can be realised by driving the cavity with a common laser field and placing detector $a$ into an interference maxima. Eq.~(\ref{conditions}) can be adjusted by maximizing the mean length of the light and dark periods and the intensity of the emitted light within a light period. Notice that the quantum dots do not have to have the same frequencies and coupling constants. Moreover, the generalisation of the scheme to the build up of large cluster states is straightforward \cite{Metz2} and opens new perspectives for scalable solid state quantum computing.

\noindent \emph{Acknowledgment.} We thank M. Atat\"ure and A. Kuhn for stimulating discussions. This work was supported in part by the Royal Society and the GCHQ, the EU RTN EMALI, the EU ToK project CAMEL, the EU IP SCALA, and the UK EPSRC through the QIP IRC.

\end{document}